\documentclass[conference]{IEEEtran}
\IEEEoverridecommandlockouts

\usepackage{cite}
\usepackage{amsmath,amssymb,amsfonts}
\usepackage{algorithm}
\usepackage{algorithmic}
\usepackage{graphicx}
\usepackage{textcomp}
\usepackage{xcolor}
\usepackage{tabularx}
\usepackage{pgfplots}
\usepackage{booktabs}
\usepackage{tikz}
\pgfplotsset{compat=1.17} 

\def\BibTeX{{\rm B\kern-.05em{\sc i\kern-.025em b}\kern-.08em
		T\kern-.1667em\lower.7ex\hbox{E}\kern-.125emX}}

\makeatletter
\newcommand{\linebreakand}{%
\end{@IEEEauthorhalign}
\hfill\mbox{}\par
\mbox{}\hfill\begin{@IEEEauthorhalign}
}
\makeatother

\begin{document}
	
\title{KScaNN: Scalable Approximate Nearest Neighbor Search on Kunpeng}

\author{
\IEEEauthorblockN{
Oleg Senkevich\textsuperscript{1\textdagger}  Siyang Xu\textsuperscript{1\textdagger} 
Tianyi Jiang\textsuperscript{1}  Alexander Radionov\textsuperscript{1}  Jan Tabaszewski\textsuperscript{1}  Dmitriy Malyshev\textsuperscript{2} \linebreakand
 Zijian Li\textsuperscript{1\#} Daihao Xue\textsuperscript{1}  Licheng Yu\textsuperscript{1}  Weidi Zeng\textsuperscript{1}  Meiling Wang\textsuperscript{1}  Xin Yao\textsuperscript{1} \linebreakand Siyu Huang\textsuperscript{1} Gleb Neshchetkin\textsuperscript{1}  Qiuling Pan\textsuperscript{1} Yaoyao Fu\textsuperscript{1}\linebreakand} 
\IEEEauthorblockA{
\textsuperscript{1}Huawei Technologies Ltd
\textsuperscript{2}Higher School of Economics
}
\thanks{\textsuperscript{\textdagger}Oleg Senkevich and Siyang Xu contributed equally to this work.}%
\thanks{\textsuperscript{\#}Corresponding author: Zijian Li (lizijian8@huawei.com).}%
}

\maketitle

\begin{abstract}
Approximate Nearest Neighbor Search (ANNS) is a cornerstone algorithm for information retrieval, recommendation systems, and machine learning applications. While x86-based architectures have historically dominated this domain, the increasing adoption of ARM-based servers in industry presents a critical need for ANNS solutions optimized on ARM architectures. A direct translation of existing x86 ANNS algorithms to ARM platforms results in a substantial performance deficit, failing to leverage the unique capabilities of the underlying hardware. To address this challenge, we introduce KScaNN, a novel ANNS algorithm co-designed for the Kunpeng 920 ARM architecture. KScaNN embodies a holistic approach that synergizes sophisticated, data-aware algorithmic refinements with carefully designed hardware-specific optimizations. Its core contributions include: 1) novel algorithmic techniques, including a hybrid intra-cluster search strategy and an improved PQ residual calculation method, which optimize the search process at a higher level; 2) an ML-driven adaptive search module that provides dynamic, per-query tuning of search parameters, eliminating the inefficiencies of static configurations; and 3) highly optimized SIMD kernels based on standard ARM NEON and SVE instructions that maximize hardware utilization for the critical distance computation workloads. The experimental results demonstrate that KScaNN not only closes the performance gap but establishes a new standard, achieving up to a 1.63x speedup over the fastest x86-based solution. This work provides a definitive blueprint for achieving leadership-class performance for vector search on modern ARM architectures and underscores the paradigm-shifting potential of hardware-software co-design.	
\end{abstract}


\section{Introduction}
\label{s1}

Nearest Neighbor Search (NNS), the task of retrieving the most similar vectors to a given query from a massive dataset, is a fundamental pillar of modern data science. It serves as a building block for a diverse array of applications, ranging from information retrieval \cite{liu, tatsuno} and recommendation systems \cite{chen} to computer vision \cite{aiger}.

Formally, the \emph{$k$-Nearest Neighbors} (\emph{$k$-NN}) problem aims to identify the set of $k$ points within a dataset $X \subset \mathbb{R}^d$ that minimize the distance to a query vector $q \in \mathbb{R}^d$ under a defined metric. However, the prohibitive computational cost of performing exact search in high-dimensional spaces has necessitated the adoption of \emph{Approximate NNS} (\emph{ANNS}) \cite{huang, gong, wang, malkov, sun}. ANNS algorithms leverage specialized indexing structures to trade a negligible reduction in recall for substantial improvements in query latency and system scalability.

ANNS is pivotal in contemporary systems where semantic similarity is paramount. Complex data objects—such as text, images, or videos—are transformed into high-dimensional vector embeddings \cite{lee, christou, le}. An ANNS algorithm then performs a similarity search to locate the top-$k$ most relevant items. For instance, modern search engines utilize this paradigm to represent both queries and documents in a shared latent space, enabling semantic retrieval that significantly outperforms traditional keyword-based matching.

Existing ANNS algorithms can be broadly categorized into four main families: hashing-based, graph-based, tree-based, and partition-based methods. \emph{Hashing-based} algorithms \cite{huang, gong} utilize hash functions to map high-dimensional data points to lower-dimensional codes, partitioning the data space into buckets to accelerate retrieval. \emph{Graph-based} algorithms \cite{wang, malkov} construct a proximity graph where nodes represent data points and edges signify similarity, performing the search via graph traversal. \emph{Tree-based} algorithms, such as the KD-tree \cite{ram}, recursively partition the data space into a hierarchical structure to efficiently locate nearest neighbors. Finally, \emph{partition-based} algorithms divide the dataset into a predefined number of disjoint clusters, confining the search to the most relevant clusters for a given query.

\begin{figure*}[t!]
	\centering
	\includegraphics[scale=0.42]{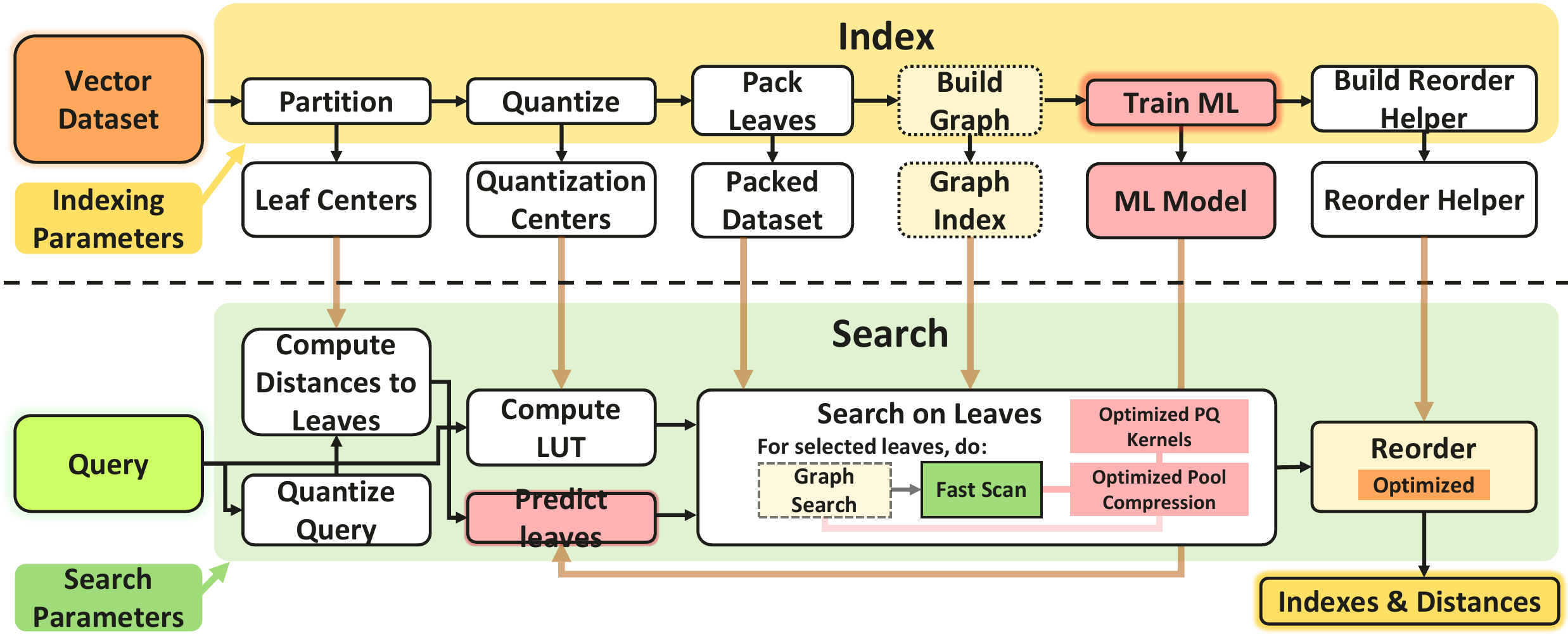}
	\caption{The pipeline of KScaNN algorithm}
	\label{fig1}
\end{figure*}

Among these families, we focus on partition-based methods, specifically the state-of-the-art Scalable Nearest Neighbors (ScaNN) algorithm \cite{guo, sun}. We selected this paradigm for two primary reasons: 1) ScaNN currently defines the performance frontier for CPU-based ANNS \cite{aumuller}, delivering superior latency-accuracy trade-offs compared to its competitors; and 2) unlike graph-based indices (e.g., HNSW \cite{malkov}), which incur substantial memory overheads, partition-based methods maintain a compact memory footprint, making them ideal for large-scale datasets in industrial applications.

While ScaNN delivers exceptional performance on x86 platforms via kernels optimized for SSE4 and AVX2/AVX512 instructions, our analysis reveals significant performance degradation when executing the algorithm on the Kunpeng 920. This disparity stems from architectural divergences and the absence of platform-specific tuning. To bridge this gap, we introduce KScaNN (Kunpeng ScaNN), a highly optimized ANNS framework that adapts and enhances ScaNN through a synergistic combination of algorithmic innovations and low-level engineering tailored for the ARM architecture.

KScaNN builds upon the core ScaNN pipeline, as illustrated in Figure \ref{fig1}. During the indexing phase, data points are partitioned into clusters and subsequently compressed using Product Quantization (PQ) \cite{pq}. At query time, a subset of candidate neighbors is rapidly identified from the compressed data and subsequently re-ranked using exact distances to yield the final top-$k$ results.

Our optimizations are developed on the Kunpeng 920 CPU, a high-performance ARM-based server processor manufactured using a 7nm process \cite{kunpeng, kunpengARM}. Despite its impressive specifications, the processor presents distinct architectural characteristics compared to x86 counterparts. Consequently, tailored low-level optimizations are required to unlock the hardware's full potential—a central focus of this work.

In summary, KScaNN represents a holistic solution that combines algorithmic refinement with hardware-aware engineering to deliver high-performance ANNS on the Kunpeng platform. Our primary contributions are as follows:

\begin{itemize}
	\item \textbf{Automated Parameter Optimization:} We introduce machine learning models to dynamically predict optimal search parameters, specifically the number of clusters to probe and the candidate pool size for re-ranking, on a per-query basis.
	\item \textbf{Advanced Algorithmic Optimizations:} We propose a predictive model for pruning empty clusters, a statistical feature-based dimensionality reduction technique, and a novel hybrid search strategy that integrates graph-based indexing within partition leaves.
	\item \textbf{ARM-Specific Kernel Optimizations:} We develop highly optimized kernels to accelerate PQ distance computations by leveraging standard ARM NEON and SVE instructions. While developed on Kunpeng, these optimizations are transferable to the broader ARM server ecosystem, including AWS Graviton \cite{graviton}.
	\item \textbf{Comprehensive Empirical Evaluation:} We conduct extensive experiments on public datasets, demonstrating that KScaNN achieves superior throughput and recall compared to state-of-the-art ANNS algorithms.
\end{itemize}

\section{Preliminaries and Background}
\label{s2}
\subsection{Notations}
\label{s2-1}

The key notations used in this paper are summarized in Table \ref{tab1}.

\begin{table}[t!]
	\caption{Summary of Notations}
	\tabcolsep=0.35cm
	\begin{center}
		\begin{tabular}{@{}cl@{}}
			\toprule
			\textbf{Notation} & \textbf{Definition} \\ \midrule
			$X$ & A dataset of vectors \\
			$x, y, x_i, \dots$ & Data vectors in $X$ \\
			$x^j$ & The $j$-th sub-vector of $x$ \\
			$n$ & The number of vectors in $X$ \\
			$d$ & The dimension of the vector space \\
			$\mathbb{R}^d$ & A $d$-dimensional real vector space \\
			$d(x, y)$ & The distance between vectors $x$ and $y$ \\ 
			$\|x\|$ & The norm of a vector $x$ \\
			$q$ & A query vector \\
			$k$ & The number of nearest neighbors to retrieve \\
			$reorder$ & The number of candidates for the re-ranking stage \\
			$L$ & The number of clusters in an IVF index \\
			$d'$ & The dimension of PQ subspaces \\
			$m$ & The number of PQ subspaces ($m = d/d'$) \\
			$K$ & The number of centroids in a $K$-means clustering \\
			$C$ & A cluster of vectors from $X$ \\
			$c$ & The centroid of a cluster $C$ \\
			$O$ & The origin of the vector space \\
			$nprob$ & The number of clusters to probe during a search \\
			$efs$ & The memory pool size for a graph-based search \\
			$X_{PQ}$ & A dataset quantized via PQ \\
			$|Y|$ & The cardinality of a set $Y$ \\
			$V, E$ & The sets of vertices and edges in a graph \\ \bottomrule
		\end{tabular}
		\label{tab1}
	\end{center}
\end{table}

\subsection{Inverted File index}
\label{s2-2}

The Inverted File (IVF) index is a widely adopted structure for efficient vector retrieval. It reduces the search space by partitioning the high-dimensional vector space into discrete regions and restricting the search to a subset of these regions. Partitioning is typically achieved via the $K$-means algorithm, which divides a dataset into $K$ distinct clusters with the objective of minimizing the intra-cluster variance.

\subsection{Product Quantization}
\label{s2-3}

Product Quantization (PQ) \cite{pq} is a vector quantization technique designed to approximate high-dimensional vectors, thereby accelerating distance computations and reducing memory costs. The PQ algorithm decomposes a $d$-dimensional vector into $m$ disjoint sub-vectors, each of dimension $d'=d/m$. A distinct codebook $\mathcal{C}^j$ is learned for each subspace $j \in \{1, \dots, m\}$ using $K$-means clustering. The training data for the $j$-th codebook consists of the set of all $j$-th sub-vectors from the original dataset. The codebooks are used to quantize the dataset. Each sub-vector is replaced by the index of its nearest centroid within the corresponding codebook, yielding a compact, code-based representation of the original high-dimensional vectors.

The efficacy of Product Quantization is intrinsically linked to the statistical structure of the data, particularly the correlation between dimensions. Fundamentally, PQ approximates the joint distribution of the high-dimensional data as the product of its marginal distributions over the subspaces. This approximation is most accurate when the subspaces are statistically independent. However, standard PQ partitions the input vector based strictly on the sequential order of features. Therefore, the quantization performance is sensitive to the dimension ordering. To minimize quantization error, the partitioning strategy should encapsulate highly correlated dimensions within the same subspace, where the local codebook can effectively model the dependency, while ensuring that the distributions across different subspaces remain as statistically independent as possible.

\subsection{Asymmetric Distance Computation}
\label{s2-4}

During the retrieval phase, asymmetric distance computation is employed to estimate the distance between an uncompressed query vector $q$ and a quantized database vector $x$:
\begin{equation}
	d(q,x) \approx \sum_{j=1}^{m} d(q^j, c_x^j),
\end{equation}
where $q^j$ denotes the $j$-th sub-vector of the query, and $c_x^j$ is the centroid associated with the $j$-th sub-vector of $x$. 

Since each subspace contains a finite number of centroids ($K$), the distances $d(q^j, c_i^j)$ for all $j \in \{1, \dots, m\}$ and $i \in \{1, \dots, K\}$ can be pre-computed once the query $q$ is received. These values are populated into $m$ lookup tables. Consequently, the approximate distance $d(q,x)$ is computed via simple table lookups and summation, avoiding the computationally expensive operations required for direct Euclidean distance calculation in high-dimensional space.

\subsection{SIMD-based PQ Optimization}
\label{s2-5}

While lookup tables significantly reduce the computational cost compared to exact distance calculation, performance can be further amplified by leveraging SIMD parallelism \cite{matsui, lut16simd}. This optimization strategy involves loading lookup tables into SIMD registers and executing lookups via SIMD shuffle instructions, thereby mitigating the latency associated with frequent memory accesses and increasing instruction throughput.

For instance, if LUT distances are quantized to 8-bit unsigned integers and the codebook size is $K=16$, a single LUT requires $16 \times 8 = 128$ bits. A 256-bit SIMD register can therefore accommodate two such LUTs, enabling the processing of two subspaces in a single instruction cycle. On Kunpeng processors, which natively utilize 128-bit NEON registers, we adopt the strategy proposed in \cite{matsui} of logically concatenating two 128-bit registers to emulate wider vector operations, ensuring that the architecture's throughput capabilities are fully utilized for asymmetric distance computation.

\begin{table}[tbp]
	\caption{Hardware specifications for Huawei Kunpeng 920 CPU and representative x86 CPUs}
	\tabcolsep=0.3cm
	\begin{center}
		\begin{tabular}{@{}cccc@{}}
			\toprule
			\textbf{Specification} & \textbf{Intel 8558P} & \textbf{AMD 9654} & \textbf{Kunpeng 920} \\ \midrule
			Cores & 48 & 96 & 80 \\
			Threads & 96 & 192 & 160 \\
			Frequency & 2.7 GHz & 2.4 GHz & 2.9 GHz \\ \bottomrule
		\end{tabular}
		\label{tab2}
	\end{center}
\end{table}

\subsection{Kunpeng Hardware Specifications}
\label{s2-6}

The Kunpeng 920 is a leading-edge server-grade ARM CPU, developed by Huawei. Fabricated using a 7nm process, it integrates up to 80 cores operating at frequencies up to 2.9 GHz. Key specifications are provided in Table \ref{tab2} for a comparison with contemporary x86 CPUs.

\section{Motivation}
\label{s3}

\subsection{Limitations of Static Configuration}
\label{s3-1}

Achieving high recall in ANNS typically requires retrieving the vast majority of true nearest neighbors. However, due to the inherent geometric complexities of high-dimensional clustering, true neighbors of a query often reside in adjacent clusters rather than the one closest to the query centroid. Consequently, the search algorithm must probe a sufficient number of clusters (\textit{nprob}) to guarantee high recall.

\begin{figure}[t!]
	\centerline{\includegraphics[scale=0.5]{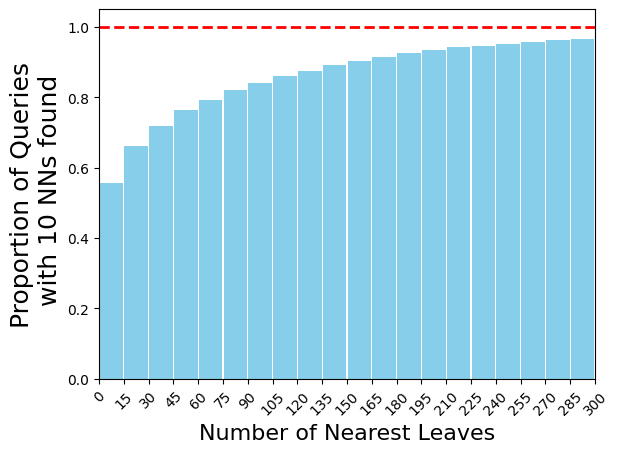}}
	\caption{Proportion of queries with 10-NN found for recall=0.99 on SIFT1M}
	\label{fig2}
\end{figure}

Crucially, the optimal \textit{nprob} is not uniform. It varies based on the query's position relative to partition boundaries. As illustrated in Figure \ref{fig2}, a static configuration imposes a rigid and often suboptimal trade-off between latency and accuracy. While \emph{hard} queries near boundaries require a large \textit{nprob} for effective retrieval, applying this configuration globally incurs significant computational redundancy for \emph{easy} queries.

\subsection{Redundancy in Cluster Probing}
\label{s3-2}

\begin{figure}[t!]
	\centerline{\includegraphics[scale=0.5]{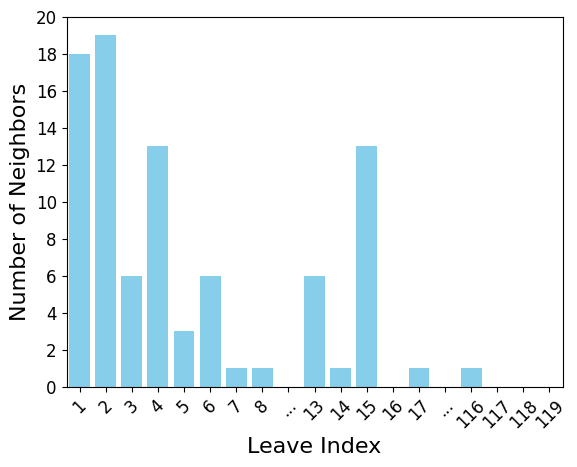}}
	\caption{Distribution of the top-100 nearest neighbors across the most proximate leaf clusters for a sample query on the GLOVE-100 dataset. Each column represents a probed cluster, revealing that the majority do not contribute to the final result set.}
	\label{fig21}
\end{figure}

A fundamental inefficiency in partition-based search arises from the heuristic nature of cluster selection. Clusters are typically prioritized based on the distance between the query and the cluster centroids. However, centroid proximity is often a suboptimal proxy for the presence of nearest neighbors. In practice, achieving high recall on million-scale datasets requires probing hundreds of leaf clusters, yet a significant proportion of these selected clusters yield no candidates among the final top-$k$ results. This discrepancy between heuristic selection and actual contribution is demonstrated in Figure \ref{fig21} using the GLOVE-100 dataset.

\subsection{Bottlenecks of Exhaustive Intra-Cluster Search}
\label{s3-3}

Once candidate clusters are identified, the standard approach in many partition-based methods is to perform an exhaustive linear scan of all data points within those clusters. Since a single cluster may contain hundreds or thousands of vectors, this brute-force approach becomes a primary latency bottleneck. To mitigate this, fine-grained data filtering and intra-cluster indexing are essential. For instance, the graph-based ANNS method KBest \cite{kbest} demonstrates that on the SIFT1M dataset, high recall is attainable with approximately 3,000 distance computations per query, merely 0.3\% of the workload required by a full scan. This disparity highlights the significant potential for performance optimization by replacing exhaustive intra-cluster scanning with more intelligent search mechanisms.

\subsection{The Necessity of Hardware-Specific Optimizations}
\label{s3-4}

Modern CPUs leverage SIMD architectures to accelerate data-parallel tasks. By utilizing wide vector registers (e.g., 128-bit or 256-bit), CPUs can execute operations on multiple data points in a single cycle, a capability critical for the distance computations that dominate ANNS workloads. However, generic compiler auto-vectorization often fails to fully exploit the specialized capabilities of the underlying hardware, such as the ARM NEON and SVE instruction sets. To maximize throughput and minimize latency, it is imperative to implement hand-tuned, SIMD-aware kernels co-designed with the hardware architecture. This ensures optimal register utilization and instruction pipelining, avoiding the overhead associated with generic implementations.

\subsection{Potential for Lightweight Dimensionality Reduction}
\label{s3-5}

The cost of vector distance computation scales linearly with data dimensionality, which accounts for over 90\% of the total query latency in many ANNS scenarios. Therefore, dimensionality reduction techniques that preserve relative distance ordering can yield substantial speedups. While standard methods like PCA are effective at reducing dimensions, they introduce additional latency during query processing by requiring the projection of the query vector into the new subspace. This limitation motivates the exploration of lightweight, feature-based dimensionality reduction techniques that reduce the computational budget without incurring significant runtime overhead or accuracy degradation.

\section{The KScaNN Algorithm}
\label{s4}

\subsection{The Algorithm Pipeline}
\label{s4-1}

KScaNN is structured into two primary stages: index construction and search. The index construction stage processes a dataset to build a specialized index structure, which the search stage then leverages to efficiently retrieve the approximate $k$-nearest neighbors for a given query vector.

During index construction, KScaNN partitions the dataset using an IVF structure and applies Product Quantization (PQ) to the residual vectors within each partition. The quantized data is reorganized into memory-aligned blocks, explicitly optimized for SIMD-based distance computation. Concurrently, a suite of machine learning models is trained to facilitate adaptive runtime optimizations during the search phase.

The search stage initiates with the computation of a Lookup Table (LUT) containing the asymmetric distances between the query vector and all PQ codebook centroids. An initial set of candidate leaf clusters is identified based on centroid proximity. This set is subsequently refined by a learned model that predicts a query-specific probing depth (\textit{nprob}). A second model dynamically tunes the size of the candidate pool required for the final re-ranking stage (\textit{reorder}).

The search then proceeds through the refined set of leaf clusters. Before processing each leaf, an optional binary classification model can be invoked to predict whether the cluster is likely to contain any nearest neighbors, allowing irrelevant clusters to be pruned. Within each leaf, KScaNN employs a hybrid search strategy: a graph-based index is used for an initial fast scan, which can transition to a brute-force search if a sufficient number of promising candidates are found. All distance computations are executed using our highly optimized SIMD kernels. Finally, the exact distances are computed for the retrieved \textit{reorder} candidates, and the top-$k$ results are returned. This process is formally described in Algorithms \ref{alg_kscann_index}, \ref{alg_kscann_search}, and \ref{alg_kscann_searchincluster}.

\begin{algorithm}[t!]
	\caption{KScaNN Index Construction}
	\label{alg_kscann_index}
	\begin{algorithmic}[1]
		\REQUIRE Dataset $X$, neighbors $k$, IVF clusters $L$, PQ centroids $K$, PQ subspace dimension $d'$
		\ENSURE Index $I$
		\IF{component filtration is enabled}
		\STATE $X_{init} \gets X$
		\STATE $X \gets \text{FilterComponents}(X)$
		\ENDIF	
		\STATE $\mathcal{C} \gets \text{KMeans}(X, L)$ \COMMENT{IVF Clustering}
		\STATE $\mathcal{PQ} \gets \emptyset$
		\FOR{each subspace $S$}
		\STATE $\mathcal{PQ} \gets \mathcal{PQ} \cup \text{KMeans}(S, K)$ \COMMENT{PQ Codebooks}
		\ENDFOR
		\STATE $X_{PQ} \gets \text{Quantize}(X, \mathcal{PQ})$
		\STATE $\mathcal{B} \gets \emptyset$ \COMMENT{SIMD Data Blocks}
		\FOR{each cluster $C \in \mathcal{C}$}
		\STATE $\mathcal{B} \gets \mathcal{B} \cup \text{BuildBlocks}(X_{PQ}, C)$
		\ENDFOR
		\STATE $\mathcal{G} \gets \emptyset$ \COMMENT{Intra-Cluster Graphs}
		\FOR{each cluster $C \in \mathcal{C}$}
		\STATE $\mathcal{G} \gets \mathcal{G} \cup \text{BuildGraph}(X, C)$
		\ENDFOR		
		\STATE $f_{\text{nprob}} \gets \text{TrainNprobPredictor}()$
		\STATE $f_{\text{reorder}} \gets \text{TrainReorderPredictor}()$
		\STATE $f_{\text{prune}} \gets \text{TrainClusterPruningPredictor}()$
		\STATE $I \gets \{X, (X_{init}), \mathcal{C}, \mathcal{PQ}, X_{PQ}, \mathcal{B}, \mathcal{G}, f_{\text{nprob}}, f_{\text{reorder}}, f_{\text{prune}}\}$
		\RETURN Index $I$
	\end{algorithmic}
\end{algorithm}

\begin{algorithm}[t!]
	\caption{KScaNN Search Procedure}
	\label{alg_kscann_search}
	\begin{algorithmic}[1]
		\REQUIRE Query $q$, index $I$, parameters $k$, $nprob_{init}$, $reorder_{init}$
		\ENSURE Top-$k$ approximate nearest neighbors
		\STATE $q_{init} \gets q$
		\IF{component filtration was used for $I$}
		\STATE $q \gets \text{FilterComponents}(q)$
		\ENDIF
		\STATE $LUT \gets \text{ComputeLUT}(q, I.\mathcal{PQ})$
		\STATE $C_{initial} \gets \text{TopNClusters}(q, I.\mathcal{C}, nprob_{init})$
		\STATE $nprob \gets I.f_{\text{nprob}}(q, C_{initial})$
		\STATE $reorder \gets I.f_{\text{reorder}}(q, nprob, reorder_{init})$
		\STATE $C_{adj} \gets \text{TopNClusters}(q, I.\mathcal{C}, nprob)$
		\STATE $R \gets \emptyset$
		\FOR{each cluster $C_j \in C_{adj}$}
		\IF{$I.f_{\text{prune}}(q, C_j) > \theta$}
		\STATE $R \gets \text{SearchInCluster}(R, I, q, C_j, LUT, reorder)$
		\ENDIF
		\ENDFOR
		\STATE Re-rank candidates in $R$, using the exact distances with $q_{init}$
		\RETURN Top-$k$ approximate nearest neighbors from $R$
	\end{algorithmic}
\end{algorithm}

\begin{algorithm}[t!]
	\caption{SearchInCluster Procedure}
	\label{alg_kscann_searchincluster}
	\begin{algorithmic}[1]
		\REQUIRE Query $q$, candidate pool $R$, index $I$, cluster $C$, LUT, $reorder$
		\ENSURE Updated candidate pool $R$
		\IF{a graph index for $C$ exists and is selected by a policy}
		\STATE $R_{graph} \gets \text{GraphSearch}(q, C, I, LUT)$
		\STATE $R \gets R \cup R_{graph}$
		\ELSE
		\FOR{each data block $b$ for $C$ in $I.\mathcal{B}$}
		\STATE $D \gets \text{CalculatePQDistancesSIMD}(LUT, b)$
		\STATE $R \gets \text{UpdateNNPoolSIMD}(R, D, reorder)$
		\ENDFOR
		\ENDIF
		\RETURN $R$
	\end{algorithmic}
\end{algorithm}

\subsection{ML-based Performance Optimization}
\label{s4-2}

\subsubsection{Adaptive Prediction of Probing Count}
\label{s4-2-1}

As discussed in Section \ref{s3-1}, a static \textit{nprob} configuration enforces an inefficient latency-accuracy trade-off. To address this, we introduce a predictive model that dynamically determines the optimal number of clusters to probe for each query. We engineered a set of lightweight features, summarized in Table \ref{tab3}, designed to capture salient query and cluster characteristics.

\begin{table*}[t!]
	\caption{Features used for $nprob$ Prediction}
	\tabcolsep=0.6cm
	\begin{center}
		\begin{tabular}{@{}llccc@{}}
			\toprule
			\textbf{Feature}  & \textbf{Description} & \textbf{Type} & \textbf{Importance} & \textbf{Cost} \\ \midrule
			Query Vector & The raw query vector components & Query Index & Low & Low \\
			Centroid Distances & The distances from a query to the nearest centroids & Query Search & High & Low \\
			Cluster Sizes & The cardinality of the nearest clusters & Cluster Index & Medium & Low \\
			Cluster Radius & The maximum intra-cluster distances from centroids & Cluster Index & Low & Medium \\
			Bayesian Scores & The pre-computed scores, indicating a cluster's quality & Cluster Search & Low & Medium \\
			Intermediate Distances & The distances of candidates, found in early stages & Search Results & High & High \\ \bottomrule
		\end{tabular}
		\label{tab3}
	\end{center}
\end{table*}

We employ a gradient boosting decision tree model \cite{lightgbm, lightgbmdoc} trained to classify queries as either \emph{easy} or \emph{hard}. The model's probabilistic output is utilized to interpolate between a minimum ($nprob_{\min}$) and maximum ($nprob_{\max}$) probing depth, resulting in a query-adaptive search scope defined as:
\begin{equation}
	\begin{aligned}
		nprob = \min(nprob_{\max}, nprob_{\min} \\ +\Delta_{nprob} \cdot \text{ReLU}(p - p_0)(p + p_1)),
	\end{aligned}
\end{equation}
where $\Delta_{nprob} = nprob_{\max} - nprob_{\min}$, $p$ denotes the model's output probability, and $p_0, p_1$ are tunable hyperparameters.

\subsubsection{Dynamic Prediction of Re-ranking Candidate Size}
\label{s4-2-2}

A similar ML-based approach is applied to dynamically predict the optimal number of candidates, \textit{reorder}, required for the final re-ranking stage. Using the same features and model architecture, we adjust the candidate pool size based on query difficulty, thereby avoiding the over-fetching of candidates for simple queries while ensuring a sufficient number for complex ones to maintain high recall. The formula for adjusting \textit{reorder} is analogous to that for \textit{nprob}.

\subsubsection{Predictive Pruning of Non-Essential Clusters}
\label{s4-2-3}

\begin{figure}[b!]
	\centerline{\includegraphics[scale=0.7]{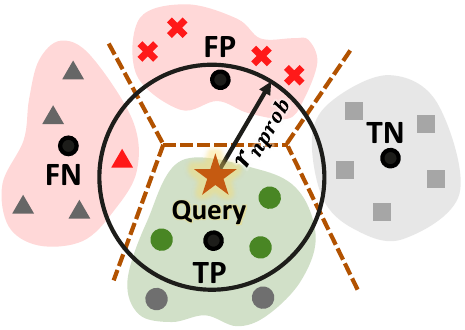}}
	\caption{Illustration of cluster pruning. FP: A cluster whose centroid is close to the query  while its points stretch in an orthogonal direction, which is unlikely to contain nearest neighbors. TP: A cluster with a relatively distant entroid that contains outlier points which are true nearest neighbors to the query.}
	\label{fig3}
\end{figure}

To further minimize computational redundancy, we introduce a binary classification model that predicts whether a selected leaf cluster is likely to contain true nearest neighbors. This mechanism enables the pruning of clusters that, while proximate in terms of centroid distance, are unlikely to contribute to the final result set. The model utilizes a rich feature set capturing the geometric and statistical properties of each cluster relative to the query, as detailed in Table \ref{tab4}. These features include relative distances, cluster cardinality, skewness (measured via inner products with principal components), and outlier statistics. These metrics facilitate the identification of clusters that may appear distant by centroid but contain relevant outliers (see Figure \ref{fig3}).

\begin{table}[t!]
	\caption{Features for Predictive Cluster Pruning}
	\begin{center}
		\begin{tabular}{@{}ll@{}}
			\toprule
			\textbf{Feature} & \textbf{Description} \\ \midrule
			$d(q, c)$ & The absolute distance from a query to a centroid \\
			relative distances & $d(q,c)$ normalized by centroid distances \\
			$|C|$ & The size of a cluster $C$ \\
			$\langle \text{pc}_i, \vec{qc} \rangle$ & The cluster skewness vs. query direction \\
			Outlier Count & Number of points in distance distribution's tail \\ 
			$r = \text{radius}(C)$ & The maximum intra-cluster distance \\
			Distance histogram & The distribution of point distances from centroids \\
			Outlier Direction & The inner product of outlier vectors with $\vec{qc}$ \\ \bottomrule
		\end{tabular}
		\label{tab4}
	\end{center}
\end{table}

\subsection{Statistical Feature-Based Dimensionality Reduction}
\label{s4-3}

The cost of distance calculations, which dominates query latency in ANNS, scales linearly with vector dimensionality. To mitigate this, we introduce an efficient, data-driven dimensionality reduction technique. This method is particularly effective for datasets where certain dimensions exhibit low variance or negligible information content, such as the uniform background regions in the FASHION-MNIST image datasets.

The method operates as a pre-processing step with zero query-time overhead, unlike transformation-based approaches like PCA which require projecting the query vector at search time. For each dimension across the entire dataset, we compute a statistical measure of its information content. Specifically, we calculate the percentage of vector components that are either zero or fall within one standard deviation of the dimension's mean. Dimensions exceeding a predefined sparsity threshold are deemed uninformative and are pruned from both the dataset and all subsequent query vectors.

This simple but effective method significantly reduces the search space dimensionality without a discernible impact on accuracy. For example, on the FASHION-MNIST dataset, this approach eliminates over 120 of the original 784 dimensions, which directly lowers the cost of distance computations and increases throughput while maintaining target recall levels.

\begin{figure}[b!]
	\centerline{\includegraphics[scale=0.9]{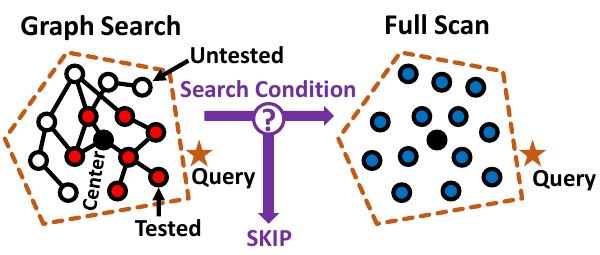}}
	\caption{The hybrid search strategy, which adaptively combines graph-based and brute-force searches within clusters based on query proximity and candidate density.}
	\label{fig5}
\end{figure}

\subsection{Hybrid Search with Intra-Cluster Graph Indexing}
\label{s4-4}

To address the bottleneck of brute-force scanning within large leaf clusters, we propose a novel hybrid search strategy. This approach integrates a leading graph index KBest \cite{kbest} to reduce distance computations while preserving high recall. A primary challenge resides in reconciling the random memory access patterns of graph traversal with our SIMD-optimized pipeline, which is designed for sequential, block-based data processing. To maintain SIMD efficiency, we structure the graph's adjacency lists into 32-point blocks.

To balance the trade-offs between graph traversal speed and brute-force thoroughness, we employ an adaptive strategy depicted in Figure \ref{fig5}. The algorithm dynamically selects the optimal method based on the cluster's proximity to the query:
\begin{itemize}
	\item For the few clusters closest to the query—which typically exhibit the highest density of true nearest neighbors—we perform an optimized brute-force scan to ensure maximum recall.
	\item For more distant clusters, we initiate a graph-based search with a restricted candidate pool size (\textit{efs}), facilitating rapid, sparse exploration.
	\item If the initial graph search identifies a sufficient number of promising candidates, the algorithm seamlessly transitions to a full brute-force scan of that cluster. This ensures that clusters containing unexpected pockets of relevant neighbors are not overlooked.
\end{itemize}

\subsection{SIMD Kernel Optimizations for ARM Architectures}
\label{s4-5}

Achieving state-of-the-art ANNS performance requires maximizing the utilization of hardware parallelism. Inspired by \cite{matsui} and \cite{lut16simd}, we developed a suite of highly optimized kernels for PQ distance computation. While fine-tuned for the Kunpeng 920 architecture, these kernels rely strictly on standard ARM NEON and Scalable Vector Extension (SVE) instruction sets, ensuring portability across the ARM ecosystem.

\subsubsection{NEON-based 2-LUT Processing}
\label{s4-5-1}

\begin{figure}[t!]
	\centerline{\includegraphics[scale=0.7]{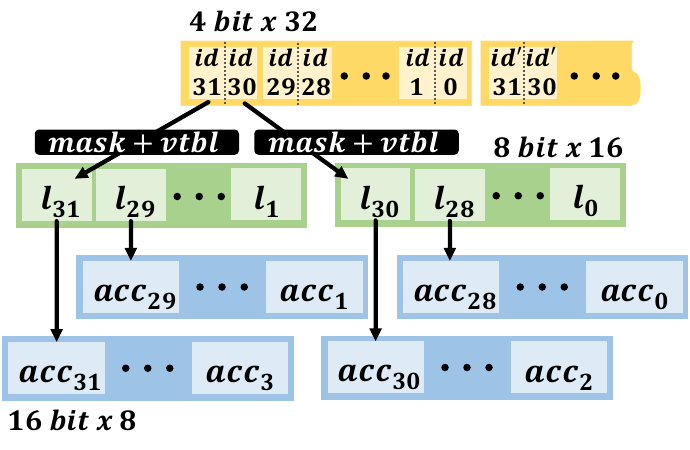}}
	\caption{The original data layout for the NEON LUT16 distance computation.}
	\label{fig6}
\end{figure}

\begin{figure}[t!]
	\centerline{\includegraphics[scale=0.72]{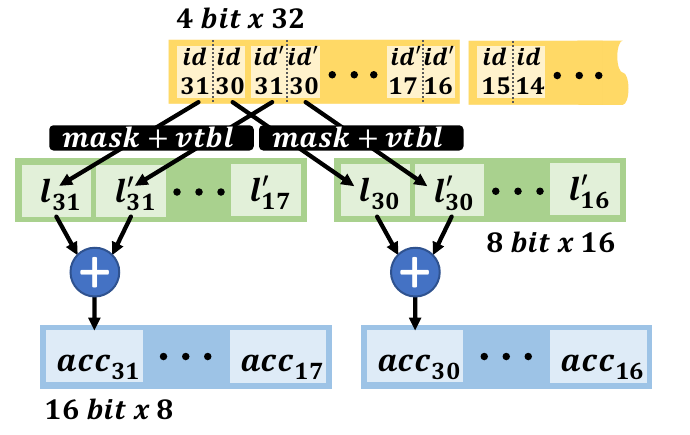}}
	\caption{The optimized data layout for the NEON LUT16 distance computation.}
	\label{fig7}
\end{figure}

To accelerate the LUT16 distance calculation on standard NEON units, we developed a kernel that circumvents the 128-bit register width limitation. By redesigning the data layout for SIMD-friendly access (Figures \ref{fig6} and \ref{fig7}), the kernel logically concatenates two 128-bit NEON registers to emulate a single 256-bit register. This enables the simultaneous processing of two LUTs, allowing two 128-bit table lookup operations to be executed in a single logical step. This technique halves the number of shuffle and accumulation instructions required per distance calculation, significantly boosting throughput.

Specifically, this kernel efficiently computes the asymmetric squared Euclidean distances for a block of points. The approximate distance for a point is the sum of its pre-computed squared sub-vector distances, which are retrieved from a series of Lookup Tables (LUTs)—one for each PQ subspace. This summation is parallelized using SIMD instructions, primarily the in-register shuffle (e.g., {TBL} in ARM NEON), which performs multiple lookups simultaneously. A key limitation of the 128-bit NEON architecture is that a single shuffle instruction can only access one 128-bit table, which can hold the 16 8-bit quantized distances for just one subspace. To circumvent this, our kernel logically concatenates two 128-bit registers to emulate a single 256-bit virtual register, thereby holding the LUTs for two distinct subspaces at once.

The kernel then involves two sequential 128-bit lookup operations: one on the lower half of this virtual register (the first LUT) and one on the upper half (the second LUT). By accumulating the results, the kernel processes the distance contributions from two subspaces in parallel, effectively doubling the lookup throughput compared to a naive, one-subspace-at-a-time approach.

This procedure operates on the 4-bit PQ codes for a block of 32 points and a global LUT containing the pre-computed squared distances from the query's sub-vectors to all centroids. The global LUT is structured as: $$d_{0,0}, d_{0,1}, \ldots, d_{0,15}, d_{1,0}, \ldots, d_{1,15}, \ldots, d_{m-1,0}, \dots, d_{m-1,15},$$ 
where each $d_{i,j}$ is an 8-bit quantized squared distance for the $j$-th centroid in the $i$-th subspace, and $m = d / d'$. The final output is a vector of accumulated squared distances for the 32 input points.

\subsubsection{NEON-based 4-LUTs processing}
\label{s4-5-2}
Extending the concept above, we implemented a more aggressive kernel that emulates a 512-bit register by concatenating four 128-bit NEON registers. This allows for the simultaneous processing of four LUTs, further improving theoretical throughput. However, this approach increases register pressure and depends heavily on the CPU's ability to manage instruction-level parallelism, making its real-world benefit architecture-dependent.

\begin{figure}[t!]
	\centerline{\includegraphics[scale=0.6]{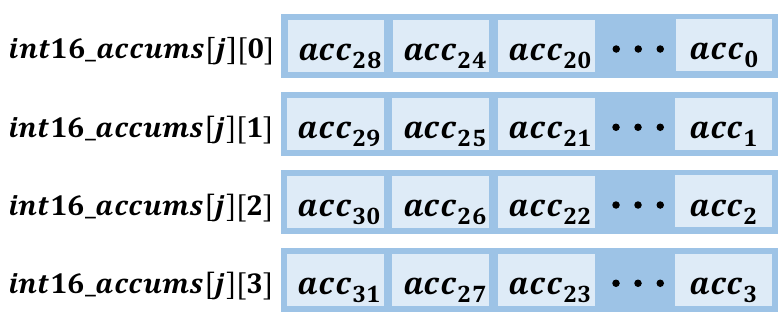}}
	\caption{The optimized data layout of computed distances for efficient SIMD-based candidate pool updates.}
	\label{fig8}
\end{figure}

\subsubsection{SVE-based 4-LUT Processing}
\label{s4-5-3}
Leveraging the Scalable Vector Extension (SVE) available on Kunpeng ARM CPUs, we developed a more advanced kernel that utilizes 256-bit SIMD registers to accelerate the distance calculations. This procedure computes the approximate squared Euclidean distances by summing pre-calculated, 8-bit quantized distances from a series of Lookup Tables (LUTs).

The core of this approach is the emulation of a single 512-bit virtual register, which is formed by logically concatenating two 256-bit SVE registers. This virtual register is large enough to hold the distance LUTs for four distinct subspaces simultaneously (16 centroids $\times$ 8-bit distance $\times$ 4 subspaces = 512 bits). The distance calculation then proceeds by executing two consecutive 256-bit table lookup instructions: one targeting the lower 256 bits of the virtual register (the first two LUTs) and another targeting the upper 256 bits (the subsequent two LUTs). By accumulating the results from these lookups, the kernel effectively processes four subspaces in a single pass, quadrupling the lookup throughput compared to a baseline 128-bit implementation.

This kernel is designed to process a specific input data layout for a block of 32 vectors. The 4-bit PQ codes are interleaved such that each byte contains the codes for two different vectors corresponding to the same subspace. This layout is structured as follows: $$v_{1,0}v_{0,0}, v_{1,1}v_{0,1}, v_{3,0}v_{2,0}, v_{3,1}v_{2,1},\ldots, v_{31,0}v_{30,0}, v_{31,1}v_{30,1},$$ where $v_{i,j}$ represents the 4-bit PQ code for vector $i$ in subspace $j$. For instance, the first byte packs the codes for vector 0 and vector 1, both from subspace 0.

The second input is the global LUT, which contains the pre-computed squared distances from the query's sub-vectors to all centroids. It is arranged contiguously by subspace: $$d_{0,0}, d_{0,1}, \dots, d_{0,15}, d_{1,0}, \dots, d_{1,15}, \dots, d_{m-1,0}, \dots, d_{m-1,15},$$ where each $d_{i,j}$ is an 8-bit quantized value, and $m = d / d'$ is the number of subspaces. The kernel outputs the final accumulated squared distances for all 32 input vectors.

\subsubsection{SVE-based Nearest Neighbors Pool Enhancement}
\label{s4-5-4}

A critical but often overlooked bottleneck is the merging of newly computed distances into the sorted candidate pool. To maintain SIMD parallelism, we designed a specialized SVE-based procedure. After computing distances for a block of 32 points, results are stored in a reordered layout (Figure \ref{fig8}). The SVE kernel loads these distances alongside the current worst candidates from the pool, performing comparison and update operations with minimal instruction overhead. This prevents costly transitions to scalar execution and ensures the entire search pipeline remains fully parallelized.

\begin{figure}[t!]
	\centerline{\includegraphics[scale=0.6]{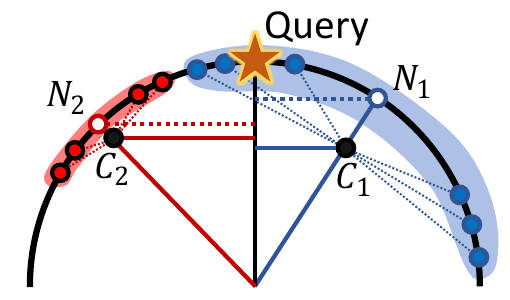}}
	\caption{Comparison of the residual vectors, computed with respect to a normalized centroid (NP) versus a non-normalized centroid (MP).}
	\label{fig9}
\end{figure}

\subsection{Additional Algorithmic Refinements}
\label{s4-6}

\subsubsection{Use of Non-Normalized Centroids for PQ Residuals}
\label{s4-6-1}

In search scenarios utilizing angular or cosine similarity, data vectors are typically normalized to the unit hypersphere. While IVF centroids are similarly normalized for initial cluster selection, we experimentally determined that using \emph{non-normalized} centroids (i.e., the arithmetic mean of the cluster points) for calculating Product Quantization residuals yields superior quantization accuracy. As illustrated in Figure \ref{fig9}, the non-normalized centroid better approximates the cluster's center of mass. Consequently, the residual vectors are more tightly concentrated around the origin, reducing the overall PQ approximation error.

\subsubsection{Other Refinements}
\label{s4-6-2}

Beyond the primary optimizations, we implemented several refinements to enhance robustness: (1) optional integration of the SOAR clustering method \cite{sun} for fuzzy partitioning, improving recall for boundary points; (2) a generalized candidate pool mutator optimized for Euclidean distance datasets; and (3) fine-tuning of $K$-means convergence parameters to ensure high-quality cluster formation.

\section{Experiments}
\label{s5}

We conducted a comprehensive empirical evaluation to validate the efficiency of KScaNN. Our primary objective is to benchmark the performance of KScaNN on the Kunpeng 920 ARM architecture against state-of-the-art ANNS implementations operating on a comparable high-performance x86 platform.\footnote{All experiments were conducted on servers located in China.} This section addresses two fundamental research questions:
\begin{itemize}
	\item \textbf{RQ1:} How does KScaNN compare against leading industry-standard baselines regarding the trade-off between search efficiency (throughput) and retrieval accuracy (recall)?
	\item \textbf{RQ2:} What is the specific performance contribution of each algorithmic and hardware-aware optimization integrated into the KScaNN framework?
\end{itemize}

\subsection{Experimental Settings}
\label{s5-1}

\textbf{Hardware Platforms.} We designed a cross-architecture comparison to demonstrate the competitiveness of KScaNN. The ARM platform features the Huawei Kunpeng 920, a flagship server-grade ARM processor. For the x86 baseline, we selected the AMD EPYC 9654, a top-tier CPU renowned for its high core count and superior single-threaded performance. This ensures that our baselines are evaluated on a leading hardware foundation. Detailed specifications are provided in Table \ref{tab_hardware}. Evaluation on additional ARM micro-architectures is reserved for future work.

\begin{table}[t!]
	\caption{Hardware Specifications of Experimental Environments}
	\tabcolsep=0.3cm
	\begin{center}
		\begin{tabular}{@{}lll@{}}
			\toprule
			\textbf{Specification} & \textbf{ARM Platform} & \textbf{x86 Platform} \\ \midrule
			Processor & 2 x Huawei Kunpeng 920 & 2 x AMD EPYC 9654 \\
			CPU Frequency & 2.9 GHz & 3.7 GHz (boost) \\
			Memory & 512 GB & 512 GB \\
			Operating System & OpenEuler 22.04 & Ubuntu 22.04 \\
			Compiler & GCC 12.3 & GCC 12.2 / clang 16.0 \\ \bottomrule
		\end{tabular}
		\label{tab_hardware}
	\end{center}
\end{table}

\begin{table}[t!]
	\caption{Datasets used for evaluation and ablation studies.}
	\tabcolsep=0.4cm
	\centering
	\begin{tabular}{@{}lrcl@{}}
		\toprule
		\textbf{Dataset}      & \textbf{Size} & \textbf{Dim.} & \textbf{Distance} \\ \midrule
		GIST \cite{gist}                  & 1M            & 960           & L2                \\
		DEEP10M \cite{deep}               & 10M           & 96            & Angular           \\
		TEXT-TO-IMAGE\cite{text2image}    & 10M           & 200           & Inner Product     \\
		BIGANN-100M \cite{bigann}         & 100M          & 128           & L2                \\
		MNIST\cite{mnist}                 & 60K           & 784           & L2                \\
		FASHION-MNIST\cite{fashionmnist}  & 60K           & 784           & L2                \\
		SIFT1M\cite{sift}                 & 1M            & 128           & L2                \\
		GLOVE-100\cite{glove}             & 1M            & 100           & Angular           \\ \bottomrule
	\end{tabular}
	\label{tab5}
\end{table}

\textbf{Evaluation Protocol.} To ensure a rigorous comparison, we adopted a strict evaluation protocol designed to measure maximum system throughput under conditions simulating a real-world online serving environment.
\begin{itemize}
	\item \textbf{Batch Size:} All search operations are performed with a batch size of one. This models typical low-latency inference scenarios where queries arrive individually and require immediate processing.
	\item \textbf{Concurrency:} To maximize hardware utilization and measure peak system throughput, we leverage all available physical CPU threads. Each thread processes an independent query stream in parallel, simulating a heavily loaded production server.
	\item \textbf{NUMA Affinity:} To eliminate cross-socket memory access latency and ensure reproducibility, each search process is pinned to a specific NUMA node. Reported throughput is the aggregate performance across all NUMA-pinned processes.
\end{itemize}

\begin{table*}[t!]
	\caption{Throughput (QPS in thousands) comparison at Recall@10 = 0.99. KScaNN is run on Kunpeng 920, while baselines are run on AMD EPYC 9654. The best baseline result for each dataset is \underline{underlined}. The \emph{Speedup} column indicates the performance gain of KScaNN over the best baseline.}
	\tabcolsep=0.58cm
	\begin{center}
		\begin{tabular}{@{}lrrrrr@{}}
			\toprule
			\textbf{Dataset} & \textbf{Google ScaNN} & \textbf{Faiss IVFPQ} & \textbf{Faiss IVFPQFastScan} & \textbf{KScaNN (Ours)} & \textbf{Speedup} \\ \midrule 
			GIST             & 38K                   & 40K                  & \underline{56K}              & \textbf{91K}           & 1.63x            \\
			DEEP10M          & 147K                  & 18K                  & \underline{212K}             & \textbf{227K}          & 1.07x            \\
			GLOVE-100        & \underline{339K}      & 4K                   & 210K                         & \textbf{360K}          & 1.06x            \\
			FASHION-MNIST    & \underline{2082K}     & 1668K                & 1687K                        & \textbf{2542K}         & 1.22x            \\ \bottomrule
		\end{tabular}
	\end{center}
	\label{tab6}
\end{table*}

\subsection{Datasets}
\label{s5-2}

Our evaluation employs a diverse suite of public benchmark datasets, summarized in Table \ref{tab5}.\footnote{For GLOVE-100 and DEEP, we randomly sampled 1M and 10M subsets, respectively, to represent common million-scale search scenarios.} These datasets were selected to span a wide range of cardinalities (60K to 100M), dimensionalities (96 to 960), and distance metrics ($L_2$, Angular, Inner Product). This variety ensures a robust assessment of our algorithm's generalizability across different data modalities and problem scales. For instance, GIST \cite{gist} and DEEP10M \cite{deep} represent high-dimensional image descriptors, GLOVE-100 \cite{glove} consists of natural language word embeddings, and TEXT-TO-IMAGE \cite{text2image} presents a challenging cross-modal retrieval task.

\subsection{Baseline Algorithms}
\label{s5-3}

We compare KScaNN against three state-of-the-art partition-based ANNS algorithms. These baselines were selected for their extensive optimization on x86 architectures, representing the current performance frontier for CPU-based vector search. For all baselines, we rigorously tuned index construction and search parameters to achieve optimal QPS at the target recall level.

\begin{itemize}
	\item \textbf{Google ScaNN (v1.4.1):} The official open-source implementation of ScaNN \cite{guo, scann}. As the conceptual predecessor to KScaNN, it serves as our primary baseline. It is heavily optimized for x86 platforms using Google's Highway library, which generates efficient AVX2/AVX512 SIMD instructions.
	\item \textbf{Faiss-IVFPQFastScan-4bit:} An aggressively optimized index from the Faiss library \cite{faiss}. This variant employs specialized 4-bit quantization and hand-tuned SIMD kernels for accelerated lookup table (LUT) distance calculations, representing a peak-performance implementation of the IVF-PQ paradigm on x86.
	\item \textbf{Faiss-IVFPQ-4bit:} A standard IVF-PQ implementation from Faiss \cite{faiss}, widely adopted in industry. This algorithm is well-optimized for modern x86 CPU architectures, serving as a robust reference baseline.
\end{itemize}

\subsection{Evaluation Metrics}
\label{s5-4}

We assess algorithm performance based on the fundamental trade-off between search efficiency and accuracy:

\begin{itemize}
	\item \textbf{Throughput (QPS):} Efficiency is quantified in Queries Per Second (QPS), defined as the total number of queries processed divided by the total wall-clock time. Higher QPS indicates superior throughput and lower operational latency.
	\item \textbf{Accuracy (Recall@10):} Search accuracy is measured by Recall@10. This metric represents the fraction of the true 10 nearest neighbors (determined via exhaustive search) successfully retrieved within the top-10 results. Reported recall is the average over all queries in the test set.
\end{itemize}

For all comparative experiments, we tune the search parameters of each method to achieve a fixed high-accuracy target of Recall@10 = 0.99. We then report the corresponding QPS at this iso-accuracy level. This methodology provides a standardized efficiency comparison representative of production systems where result quality is paramount.

\subsection{Overall Performance Comparison (RQ1)}
\label{s6}

Table \ref{tab6} presents the core results of our cross-architecture comparison. The findings demonstrate that KScaNN not only bridges the performance gap between ARM and x86 for this demanding workload but consistently outperforms the most advanced baselines running on a top-tier x86 server. KScaNN achieves a speedup ranging from 1.06$\times$ to 1.63$\times$ over the best baseline on each respective dataset. A detailed analysis reveals key insights into KScaNN's performance advantages:
 
\begin{figure*}[t!]
	\centering
	\begin{tikzpicture}
		\definecolor{c1}{RGB}{128,128,128} 
		\definecolor{c2}{RGB}{65,105,225}  
		\definecolor{c3}{RGB}{220,20,60}   
		\definecolor{c4}{RGB}{255,140,0}   
		\definecolor{c5}{RGB}{34,139,34}   
		\definecolor{c6}{RGB}{138,43,226}  
		\begin{axis}[
			width=\textwidth,
			height=6cm,
			ybar,
			bar width=9pt,
			ymin=0,
			enlarge x limits=0.15,
			legend style={
				at={(0.5,1.02)}, 
				anchor=south, 
				legend columns=-1, 
				draw=none, 
				fill=none, 
				font=\normalsize,
				/tikz/every odd column/.append style={column sep=3pt}, 
				/tikz/every even column/.append style={column sep=15pt} 
			},
			legend image code/.code={
				\draw[#1, draw=none] (0cm,-0.1cm) rectangle (0.6cm,0.1cm);
			},
			ylabel={Normalized QPS},
			symbolic x coords={GIST,DEEP10M,TEXT-TO-IMAGE,BIGANN,FASHION-MNIST,MNIST},
			xtick={GIST,DEEP10M,TEXT-TO-IMAGE,BIGANN,FASHION-MNIST,MNIST},
			axis line style={-},
			axis lines*=left,
			ymajorgrids=true,
			grid style={dashed, gray!30},
			nodes near coords,
			nodes near coords align={vertical},
			nodes near coords style={font=\footnotesize, rotate=90, anchor=west, color=black},
			xticklabel style={font=\small, rotate=0, align=center}, 
			yticklabel style={font=\small},      
			ylabel style={font=\small},          
			cycle list={
				{fill=c1, draw=none},
				{fill=c2, draw=none},
				{fill=c3, draw=none},
				{fill=c4, draw=none},
				{fill=c5, draw=none},
				{fill=c6, draw=none}
			}
			]
			\addplot coordinates {(GIST,56) (DEEP10M,212) (FASHION-MNIST,208)};
			\addplot coordinates {(GIST,2) (DEEP10M,5) (TEXT-TO-IMAGE,92) (FASHION-MNIST,73)};
			\addplot coordinates {(GIST,39) (DEEP10M,112) (TEXT-TO-IMAGE,94) (FASHION-MNIST,125)};
			\addplot coordinates {(GIST,73) (DEEP10M,170) (TEXT-TO-IMAGE,124) (BIGANN,17) (FASHION-MNIST,206) (MNIST, 199)};
			\addplot coordinates {(GIST,90) (DEEP10M,226) (TEXT-TO-IMAGE,130) (BIGANN,28) (FASHION-MNIST,222) (MNIST, 217)};
			\addplot coordinates {(FASHION-MNIST,254) (MNIST, 285)};
			\legend{x86, Base, +Arm, +MinorOpt, +ML, +CompFilt}
		\end{axis}
	\end{tikzpicture}
	\caption{Ablation study showing cumulative QPS improvement as KScaNN features are incrementally enabled. All results are at Recall@10=0.99. For each dataset, we normalize the QPS of all methods by a common divisor, respectively. The \emph{Base} and \emph{+Arm} versions fail to run on the BIGANN dataset, and the \emph{+CompFilt} optimization only achieves improvements on datasets FASHION-MNIST and MNIST. Vertical axis shows thousands QPS for GIST, DEEP and BIGANN, tens of thousands QPS for FASHION-MNIST and MNIST, hundreds QPS for TEXT-TO-IMAGE.}
	\label{fig91}
\end{figure*}
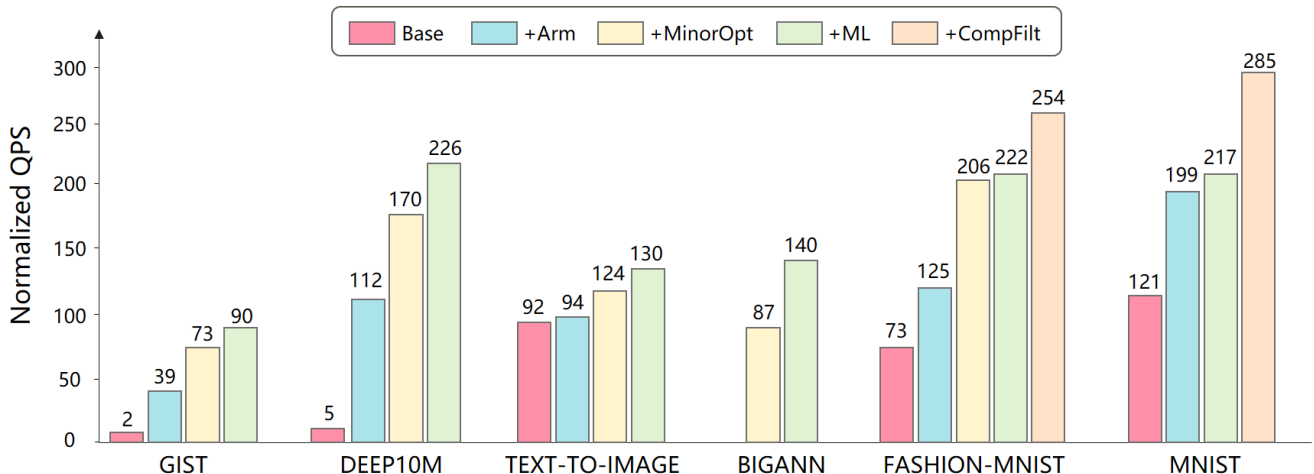

\begin{figure}[t!]
	\centerline{\includegraphics[scale=0.63]{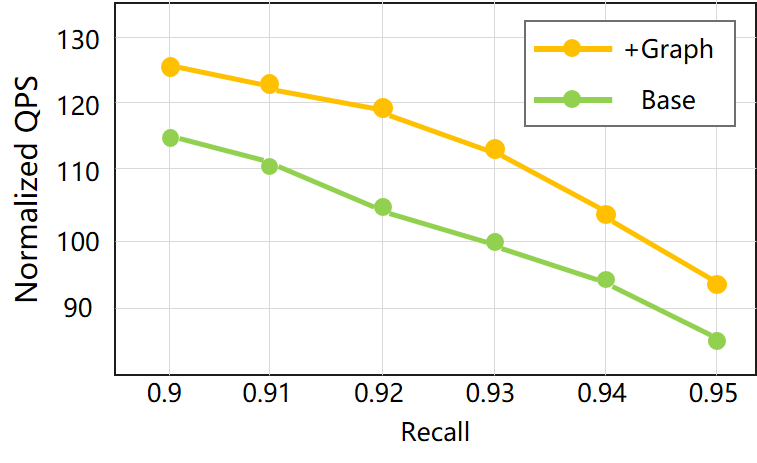}}
	\caption{Performance of the hybrid graph-based search on SIFT1M, showing consistent QPS improvement over the baseline brute-force scan within leaves. We normalize the QPS of the \emph{Base} and \emph{+Graph} versions by a common divisor.}
	\label{fig92}
\end{figure}

\begin{itemize}
	\item \textbf{Compute-Bound Workloads:} The high-dimensional GIST dataset (960 dimensions) renders the search process intensely compute-bound. Here, KScaNN achieves a remarkable \emph{1.63$\times$} speedup over the fastest baseline. This gain is a direct testament to the efficiency of our custom ARM NEON and SVE kernels, which extract higher computational throughput per clock cycle from the Kunpeng hardware than the mature AVX-optimized kernels of the baselines can from the x86 platform.
	\item \textbf{Memory-Bound Workloads:} Conversely, the lower-dimensional DEEP10M dataset (96 dimensions) shifts the bottleneck toward memory access and pipeline efficiency. Even here, KScaNN's holistic design—featuring memory-aligned data blocks and an optimized search pipeline—delivers a solid \emph{1.07$\times$} speedup, demonstrating balanced performance when raw SIMD computation is not the sole dominant factor.
	\item \textbf{Algorithmic Impact on Angular Search:} On GLOVE-100, which utilizes angular distance, KScaNN's \emph{1.06$\times$} improvement over Google ScaNN is amplified by algorithmic choices. This advantage stems from our novel use of non-normalized centroids for PQ residual calculation (Section \ref{s4-6-1}), which yields more accurate quantization for data on the hypersphere.
	\item \textbf{Data-Aware Optimization:} The most striking algorithmic improvement appears on FASHION-MNIST. This dataset contains significant redundancy, with many dimensions corresponding to uniform image backgrounds. Our statistical component filtration method (Section \ref{s4-3}) automatically identifies and prunes over 120 uninformative dimensions at zero query-time cost. This fundamentally reduces the computational workload, enabling a \emph{1.22$\times$} speedup and highlighting the efficacy of data-aware preprocessing.
\end{itemize}

\subsection{Ablation Study (RQ2)}
\label{s7}

To dissect the contributions of our optimization strategy, we conducted a thorough ablation study. We began with a \emph{Base} implementation—a direct port of the ScaNN algorithm compiled for Kunpeng without hardware-specific optimizations—and incrementally enabled each major KScaNN optimization. The cumulative performance gain at Recall@10 = 0.99 is summarized in Figure \ref{fig91}.

The study reveals a layered path to high performance. The \emph{Base} version lags substantially behind x86 baselines, confirming that a naive port is insufficient. The most significant improvement, yielding up to a 20-fold increase on the compute-bound GIST dataset, comes from our hardware-specific ARM NEON and SVE kernels (\emph{+Arm}). This underscores our central thesis: low-level, architecture-aware SIMD optimization is essential for competitive performance.

Building on this, the \emph{+MinorOpt} stage (incorporating SOAR clustering and improved data structures) provides a consistent 50\%--80\% boost. Subsequently, ML-based adaptive parameter tuning (\emph{+ML}) yields another major improvement, especially on heterogeneous datasets like BIGANN. By dynamically tailoring \emph{nprob} and \emph{reorder} for each query, the system avoids the inefficiency of a static, one-size-fits-all approach.

Furthermore, the \emph{+CompFilt} stage provides a decisive speedup, but exclusively on FASHION-MNIST and MNIST. This specificity arises from the nature of the vector representations. MNIST and FASHION-MNIST vectors are raw, flattened pixel values with large, uniform backgrounds, resulting in many low-variance dimensions that our component filtration detects and prunes. In contrast, datasets like GIST consist of manually engineered feature descriptors designed to be information-dense. These descriptors have effectively already filtered the redundancy that \emph{+CompFilt} targets.

To evaluate our hybrid graph-in-leaf strategy, we analyzed the SIFT1M dataset. As shown in Figure \ref{fig92}, enabling hybrid search (\emph{+Graph}) provides a consistent QPS improvement of 10\%--15\% across the 0.9 to 0.95 recall range compared to the baseline brute-force scan. This validates our approach of combining rapid graph-based exploration in sparse clusters with exhaustive scanning in dense ones.

\subsection{Generalizability Tests on the x86 platform}
\label{s71}

We investigated the generalizability of our algorithmic improvements on x86 by isolating them from ARM-specific SIMD optimizations. Experiments were conducted on an Intel Xeon Platinum 8378A CPU (3.00 GHz, 32 cores, 512 GB RAM). We integrated KScaNN's algorithmic enhancements into the baseline x86 implementation, replacing ARM-specific SIMD optimizations with standard x86 instructions. As illustrated in Figure \ref{fig911}, the algorithmic optimizations yield performance gains on x86 hardware as well, demonstrating that the proposed improvements are effective across architectures.

\begin{figure*}[t!]
	\centering
	\begin{tikzpicture}
		\definecolor{c2}{RGB}{65,105,225}  
		\definecolor{c4}{RGB}{255,140,0}   
		\definecolor{c5}{RGB}{34,139,34}   
		\definecolor{c6}{RGB}{138,43,226}  
		\begin{axis}[
			width=\textwidth,
			height=5.9cm,
			ybar,
			bar width=9pt,
			ymin=0,
			enlarge x limits=0.15,
			legend style={
				at={(0.5,1.02)}, 
				anchor=south, 
				legend columns=-1, 
				draw=none, 
				fill=none, 
				font=\normalsize,
				/tikz/every odd column/.append style={column sep=3pt}, 
				/tikz/every even column/.append style={column sep=15pt} 
			},
			legend image code/.code={
				\draw[#1, draw=none] (0cm,-0.1cm) rectangle (0.6cm,0.1cm);
			},
			ylabel={Normalized QPS},
			symbolic x coords={GIST,DEEP10M,TEXT-TO-IMAGE,BIGANN,FASHION-MNIST,MNIST},
			xtick={GIST,DEEP10M,TEXT-TO-IMAGE,BIGANN,FASHION-MNIST,MNIST},
			axis line style={-},
			axis lines*=left,
			ymajorgrids=true,
			grid style={dashed, gray!30},
			nodes near coords,
			nodes near coords align={vertical},
			nodes near coords style={font=\footnotesize, rotate=90, anchor=west, color=black},
			xticklabel style={font=\small, rotate=0, align=center}, 
			yticklabel style={font=\small},      
			ylabel style={font=\small},          
			cycle list={
				{fill=c2, draw=none},
				{fill=c4, draw=none},
				{fill=c5, draw=none},
				{fill=c6, draw=none}
			}
			]
			\addplot coordinates {(GIST,60) (DEEP10M,17) (TEXT-TO-IMAGE,23) (BIGANN,42) (FASHION-MNIST,120) (MNIST, 122)};
			\addplot coordinates {(GIST,61) (DEEP10M,28) (TEXT-TO-IMAGE,30) (BIGANN,42) (FASHION-MNIST,122) (MNIST, 132)};
			\addplot coordinates {(GIST,78) (DEEP10M,40) (TEXT-TO-IMAGE,32) (BIGANN,57) (FASHION-MNIST,140) (MNIST, 139)};
			\addplot coordinates {(FASHION-MNIST,149) (MNIST, 186)};
			\legend{Base, +MinorOpt, +ML, +CompFilt}
		\end{axis}
	\end{tikzpicture}
	\caption{Ablation study showing cumulative QPS improvement as KScaNN features are incrementally enabled on x86. All results are at Recall@10=0.99. For each dataset, we normalize the QPS of all methods by a common divisor, respectively. Vertical axis shows hundreds QPS for GIST, TEXT-TO-IMAGE and BIGANN, thousands QPS for DEEP, FASHION-MNIST and MNIST}
	\label{fig911}
\end{figure*}

\subsection{Impact of Data Distribution on Search Efficiency}
\label{s9}

We observe distinct interaction patterns between data distribution characteristics and KScaNN's search efficiency:
\begin{itemize}
	\item For datasets with \emph{dense feature spaces} (e.g., GIST), information is uniformly distributed, making dimensionality reduction challenging. Here, performance improvements are driven by optimized SIMD kernels, which maximize throughput while algorithmic pruning is less effective.
	\item For datasets with \emph{high feature redundancy} (e.g., FASHION-MNIST), \emph{Component Filtration} delivers the most significant speedup. By exploiting statistical sparsity, this optimization effectively prunes uninformative dimensions, directly reducing computational costs.
	\item For datasets with \emph{skewed cluster sizes}, ML-based adaptive search proves most effective. By dynamically adjusting search parameters based on local density, KScaNN mitigates latency penalties associated with ``hard'' queries, ensuring robust performance.
	\item Generally, the optimal configuration of KScaNN is intrinsically linked to dataset clustering patterns. KScaNN relies on dataset-specific tuning of leaf cluster counts and probing depth (\emph{nprob}) to align the index structure with the underlying data distribution.
\end{itemize}

\section{Discussion and Future Work}
\label{s8}

\subsection{KScaNN's Generalizability across the ARM Ecosystem}
\label{s8-0}

Although KScaNN was developed and optimized on Kunpeng 920, its underlying design principles possess intrinsic portability across the broader ARM server ecosystem such as AWS Graviton \cite{graviton}.

First, the proposed SIMD kernels rely strictly on standard ARM NEON and SVE instruction sets, ensuring binary or source-level compatibility with other ARM-based CPUs. Furthermore, by adhering to the Vector Length Agnostic (VLA) programming model in our SVE implementation, the kernels can adapt seamlessly to processors with varying vector widths (e.g., the 256-bit width of AWS Graviton 3) without requiring code modification.

Second, the algorithmic innovations within KScaNN, e.g., the ML-based parameter tuning and the hybrid graph-based search, are architecture-agnostic. These optimizations reduce total computational complexity and enhance cache locality, yielding performance benefits that are independent of the specific underlying hardware architecture.

\subsection{The Challenge of Effective Pruning on Modern Hardware}
\label{s8-1}

As part of this study, we investigated several geometrically motivated pruning strategies intended to reduce the number of distance calculations within leaf clusters. Despite their theoretical elegance and the ability to identify a significant fraction of non-candidates during offline analysis (up to 30\% in certain instances), none of these strategies yielded a net improvement in online query throughput.

This counter-intuitive outcome underscores a critical reality of modern CPU architectures: the computational cost of our highly optimized SIMD kernels is exceptionally low. However, the overhead introduced by the control logic, branching, and scalar computations required to evaluate geometric pruning conditions consistently outweighs the savings gained from avoided distance calculations. Therefore, for a pruning method to be viable in this high-throughput regime, it must either be integrated directly into the SIMD pipeline without incurring control flow divergence or be capable of eliminating a substantially larger fraction of the search space at negligible cost.

\subsection{Future Work}
\label{s8-2}

The success of KScaNN opens several avenues for future research on ARM-based platforms:
\begin{itemize}
	\item \textbf{Architecture-Specific Optimizations:} We aim to develop optimizations deeply coupled with ARM architectural features, including cache hierarchy analysis for improved graph traversal and kernels for the next-generation ARM Scalable Vector Extension 2 (SVE2).
	\item \textbf{Advanced Quantization and Dynamic Indexing:} We plan to explore quantization schemes beyond standard PQ, such as RaBitQ \cite{gao}, and extend KScaNN to support efficient, low-latency insertions and deletions—features critical for production environments.
	\item \textbf{Hybrid Indexing Structures:} Future work could integrate tree-based structures for coarse-grained partitioning or investigate multi-level graph indexes to better accommodate datasets with highly non-uniform density distributions.
	\item \textbf{Data Filtration Strategies:} We intend to revisit geometrically motivated filtration strategies (e.g., strip boundaries, convex hulls) with a focus on reducing control-flow overhead. Additionally, we will investigate partial PQ distance evaluation to prune candidates in distant leaf nodes by reordering PQ components based on standard deviation.
\end{itemize}

\section{Conclusion}
\label{s9}

We introduced KScaNN, a high-performance ANNS algorithm that demonstrates the profound efficacy of a hardware-software co-design philosophy. By combining data-aware algorithmic refinements with optimized kernels tailored for the ARM architecture, KScaNN not only bridges the historical performance gap between ARM and x86 platforms but also establishes a new standard for vector retrieval efficiency.

The key innovations of KScaNN are threefold: 1) novel algorithmic enhancements, including a hybrid intra-cluster search strategy and an improved PQ residual calculation method, which optimize the search process at a high level; 2) an ML-driven adaptive search module that provides dynamic, per-query tuning of search parameters, thereby eliminating the inefficiencies of static configurations; and 3) highly optimized SIMD kernels for ARM that maximize hardware utilization for critical distance computation workloads.

Our experiments confirm the superiority of this approach. KScaNN, running on a Kunpeng 920 CPU, consistently outperforms state-of-the-art baselines running on top-tier x86 processors, achieving a relative speedup of up to 1.63$\times$ at a high-accuracy target of 99\% recall. This work provides a definitive blueprint for high-performance vector search on modern ARM architectures and validates the broader shift towards co-design, where achieving peak performance requires the intimate and intelligent integration of software algorithms and hardware capabilities.

\end{document}